\documentclass[superscriptaddress,floatfix,twocolumn,showpacs,preprintnumbers,amsmath,amssymb,prb]{revtex4}

\usepackage[dvipdfm]{graphicx}%Include figure files
\usepackage{dcolumn}% Align table columns on decimal point
\usepackage{bm}% bold math
\begin{document}

\preprint{APS/123-QED}

\title{Magnetic Field versus Temperature Phase Diagram of the Spin-1/2 Alternating Chain Compound F$_5$PNN}% line breaks with \\
\author{Yasuo Yoshida}
\altaffiliation[Present address: ]{Department of Physics, University of Florida, PO. Box 118440, Gainesville, FL 32611-8440, USA}
\altaffiliation[E-mail: ]{yoshida@phys.ufl.edu}
\affiliation{Department of Applied Quantum Physics, Kyushu University, Fukuoka 812-8581, Japan}
\author{Tatsuya Kawae}
\affiliation{Department of Applied Quantum Physics, Kyushu University, Fukuoka 812-8581, Japan}
\author{Yuko Hosokoshi}
\affiliation{Department of Physical Science, Osaka Prefecture University, Sakai, Osaka 599-8531, Japan}
\author{Katsuya Inoue}
\affiliation{Department of Chemistry, Hiroshima University, Higashi-Hiroshima, Hiroshima 739-0024, Japan}
\author{Nobuya Maeshima}
\affiliation{Department of Chemistry, Tohoku University, Sendai, Miyagi 980-8578, Japan}
\affiliation{Institute for Molecular Science, Okazaki, Aichi 444-8585, Japan}
\author{Koichi Okunishi}
\affiliation{Department of Physics, Niigata University, Niigata 950-2181, Japan}
\author{Kiyomi Okamoto}
\affiliation{Department of Physics, Tokyo Institute of Technology, Tokyo 152-8551, Japan}
\author{Toru Sakai}
\affiliation{Japan Atomic Energy Agency, SPring-8, Sayo, Hyogo 679-5148, Japan}
%$^{1}$ Department of Physics, Kyushu Sangyo University, Matsukadai, Fukuoka 813\\
%$^{2}$ Department of Physical Science, Osaka Prefecture University, Sakai, Osaka 599-8531, Japan\\
%$^{3}$ Department of Chemistry, Hiroshima University, Higashi-Hiroshima, Hiroshima 739-0024, Japan\\
%$^{4}$ Department of Chemistry, Tohoku University, Sendai, Miyagi 980-8578, Japan\\
%$^{5}$ Institute for Molecular Science, Okazaki, Aichi 444-8585, Japan\\
%$^{6}$ Department of Physics, Niigata University, Niigata 950-2181, Japan\\
%$^{7}$ Department of Physics, Tokyo Institute of Technology, Tokyo 152-8551, Japan\\ 
%$^{8}$ Japan Atomic Energy Agency, SPring-8, Sayo, Hyogo 679-5148, Japan.}
%$^{9}$ Department of Physics, University of Florida, PO. Box 118440, Gainesville, FL 32611-8440, USA.}
\date{\today}% It is always \today, today,
             %  but any date may be explicitly specified
\begin{abstract}
We have measured the specific heat of the $S=1/2$ alternating Heisenberg antiferromagnetic chain compound 
pentafluorophenyl nitronyl nitroxide in magnetic fields % and at temperatures down to 130 mK, 
using a single crystal and powder. 
A sharp peak due to field-induced magnetic ordering (FIMO) is observed in both samples. 
%The shape of the $H$-$T$ phase boundary of the FIMO is different between the two samples.   
The $H$-$T$ phase boundary of the FIMO of the single crystal is symmetric with respect to the central field  
of the gapless field region $H_\mathrm{C1}\leq H\leq H_\mathrm{C2}$, whereas 
%as observed in the other isotropic spin-gapped compounds. 
it is distorted for the powder whose ordering temperatures are lower. 
An analysis employing calculations based on the finite temperature 
density matrix renormalization group 
indicates the possibility of novel incommensurate ordering due to frustration in the powder around the central field.  
\end{abstract}
% PACS, the Physics and Astronomy
                             % Classification Scheme.
%\keywords{Suggested keywords}%Use showkeys class option if keyword
                              %display desired
\maketitle
\section{Introduction}
Field-induced magnetic ordering (FIMO) in spin-gapped systems, in which
an energy gap exists for low-lying excited states, has been investigated
in a vast number of compounds, particularly in the context of the Bose-Einstein condensation (BEC) of triplet magnons \cite{nikuni}.
The BEC picture is useful for understanding the nature of the FIMO
with the commensurate (C) antiferromagnetic order pependicular to the field direction. 
%Field-induced magnetic ordering (FIMO) in spin-gapped systems, 
%in which an energy gap exists for low-lying excited states, 
%has been investigated in a large number of compounds 
%since Nikuni {\it et al.} proposed the Bose-Einstein condensation (BEC) of triplet magnons 
%as a description of the FIMO \cite{nikuni}. 
Recently, Suzuki {\it et al.} and Maeshima {\it et al.} have added a new aspect 
to the FIMO on the basis of numerical analyses combined with field theories \cite{suzuki, maeshima}. 
%in connection with Tomonaga-Luttinger liquid and frustration \cite{suzuki, maeshima}. 
These authors have predicted that a magnetic field induces a novel 
incommensurate (IC) order parallel to the field direction 
in $S=1/2$ alternating chains with a frustrated next-nearest-neighbor (NNN) 
interaction. 
Around the central field of the field-induced Tomonaga-Luttinger liquid (TLL) 
phase of this system between the lower and upper critical field, $H_{\rm C1}$ and $H_{\rm 
C2}$, frustration changes the dominant spin correlation from C to IC.
If small inter-chain interactions exist, 
the dominant IC correlation leads to long-range IC ordering in the field direction. 
In the case frustration is not strong enough to stabilize the IC order at high temperatures, 
a first order phase transition will happen from the BEC to the IC order at very low temperatures\cite{maeshima, maeshima2}.

The theoretical studies mentioned above \cite{suzuki, maeshima} have been stimulated by experimental works on 
the organic radical compound pentafluorophenyl nitroxide (F$_5$PNN) \cite{across, izumi, izumithesis, lt23}. 
The magnetism of F$_5$PNN arises from unpaired electrons delocalized around the NO moieties. 
Although this compound has a uniform chain structure at room temperature, 
the magnetic susceptibility and the magnetization curve at low temperatures 
are well reproduced by calculations for an $S=1/2$ alternating chain model 
which is described by the spin Hamiltonian \cite{across};  
\begin{equation} 
 H = -2J \sum_{i}^{N/2} (S_{2i-1}\cdot S_{2i}+\alpha S_{2i}\cdot S_{2i+1}). 
\end{equation}
Here, $S$ denotes the $S=1/2$ Heisenberg-type spin operator, $N$ is the total number of spins, and 
$\alpha$ is the alternation ratio between competing two nearest-neighbor interactions in a one-dimensional chain. 
When $\alpha$=1, the system becomes a uniform chain, whereas when $\alpha $=0 
the system breaks up into the assembly of isolated dimers. 
In Ref. \onlinecite{across}, 
the alternation ratio $\alpha=0.4$ and exchange interaction $2J/k_\mathrm{B}=-5.6$ K 
were obtained for F$_5$PNN.
 
The lower and upper critical fields of F$_5$PNN are determined to be about $H_\mathrm{C1}=3.0$ T and $H_\mathrm{C2}=6.5$ T 
from the magnetization curve.  
%Since these critical fields are much lower than those of any other spin-gapped compounds, 
%F$_5$PNN is suitable to investigate FIMO in spin-gapped systems.    
NMR shows a TLL behavior in spin-lattice relaxation 
and provides evidence for a NNN interaction \cite{izumi, suga, izumithesis}. 
In previous works, 
we observed FIMO by measuring the specific heat of a polycrystalline sample 
in magnetic fields up to 8.0 T ( $>H_\mathrm{C2}$ ) \cite{prl}. 
Above the critical temperature of the FIMO, 
the temperature dependence of the specific heat $C(T)$ in magnetic fields was 
in good qualitative agreement with a numerical calculation which assumes the TLL \cite{prl, wang}. 

In this paper, we present the $H$-$T$ phase diagrams of a single crystal and powder of F$_5$PNN
obtained from detailed specific heat measurements in magnetic fields. 
%One of two samples is the relatively large single crystal sample and 
%the other is small powder-like crystal sample which are obtained from the broken large single crystals. 
Reentrant $H$-$T$ phase diagrams for the FIMO phase are obtained for both samples. 
However, the shape of the phase boundary depends on the form of the sample. 
That of the single crystal is symmetric with respect to a central field of the gapless field region between $H_\mathrm{C1}$ and $H_\mathrm{C2}$, 
whereas the powder has a phase boundary which is distorted and pushed to lower temperatures than that of the single crystal. 

\section{Experimental procedures}
F$_5$PNN was prepared using the method described in Ref. \onlinecite{sample}. 
Specific heat measurements were performed by the adiabatic heat-pulse method using a $^3$He-$^4$He dilution refrigerator.
The powder sample was mixed with Apiezon N grease to ensure good thermal contact, 
and was mounted on the sample cell in the refrigerator.
The single crystal sample was attached to the cell with the same grease.  
The nuclear contributions of hydrogen and fluorine to the specific heat were subtracted.
% from the data.
%We use 50 mg of powder and 20 mg of single crystals for respective measurements. 

%\section{Results and Discussion}
\begin{figure}[t]
\begin{center}
   \includegraphics[width=8cm]{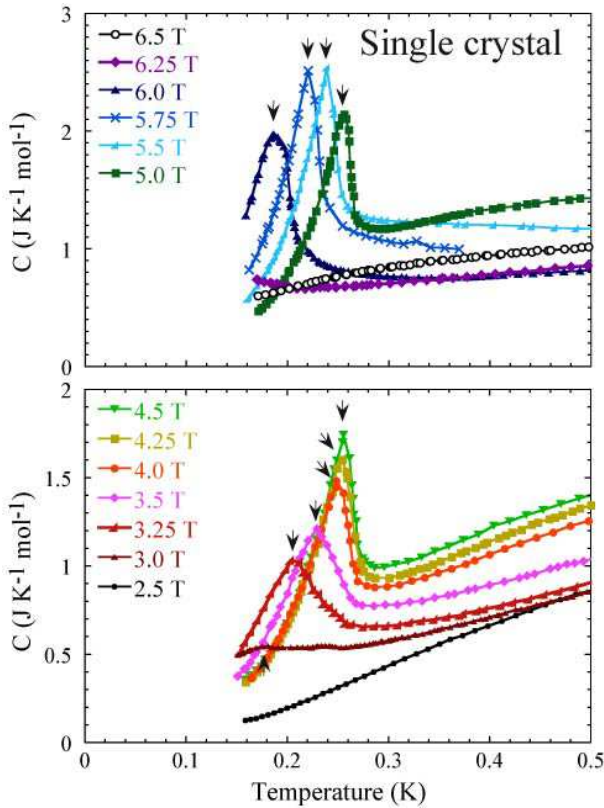}% Here is how to import EPS art
\end{center}
%\begin{center}
%   \includegraphics[width=8cm]{FIMOsinglehighHConverted.eps}% Here is how to import EPS art
%\end{center} 
\caption{(color online) Specific heat of the single crystal 
in magnetic fields. Arrows indicate peak temperatures of the FIMO. 
Upper panel: 5.0 T $\leq H\leq$ 6.5 T. 
Lower panel: 2.5 T $\leq H\leq$ 4.5 T. 
%Data in fields shown in both glaphs are shifted in the absolute value of specific heat for ease to understand the behaviors. 
%and indicating the FIMO phases for single and polycrystalline samples, respectively. 
}
\end{figure}

\begin{figure}[t]
\vspace{0.18cm}
\begin{center}
   \includegraphics[width=8.3cm]{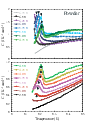}% Here is how to import EPS art
\end{center}
%\begin{center}
%   \includegraphics[width=8cm]{FIMOpowderhighHConverted.eps}% Here is how to import EPS art
%\end{center} 
\vspace{0.25cm}
\caption{(color online) Specific heat of the powder in magnetic fields. 
Arrows indicate peak temperatures of the FIMO.  
Upper panel: 4.75 T $\leq H\leq$ 6.75 T. 
Lower panel: 2.5 T $\leq H\leq$ 4.5 T. 
%Data in fields shown in both glaphs are shifted in the absolute value of specific heat for ease to understand the behaviors. 
%and indicating the FIMO phases for single and polycrystalline samples, respectively. 
}
\end{figure}

%\begin{figure}[bp]
%   \includegraphics[width=8cm]{sakai&singlepowderPD3.eps}% Here is how to import EPS art
% \caption{Magnetic field versus temperature phase diagram of single crystal and powder-like samples of F$_5$PNN obtained 
%from the specific heat measurements. Open and filled circles are peak positions of the specific heats of single crystal and 
%powder-like samples, respectively. Solid lines are guides to eyes.
%%and indicating the FIMO phases for single and polycrystalline samples, respectively. 
%}
%\end{figure}

%\begin{figure}[bp]
%   \includegraphics[width=8.3cm]{suzukiPD+f5pnn.eps}% Here is how to import EPS art
% \caption{Magnetic field versus temperature phase diagram of single crystal and powder-like samples of F$_5$PNN obtained 
%from the specific heat measurements. Open and filled circles are peak positions of the specific heats of single crystal and 
%powder-like samples, respectively. Solid lines are guides to eyes.
%%and indicating the FIMO phases for single and polycrystalline samples, respectively. 
%}
%\end{figure}

\section{Results}
Figure 1 shows $C(T)$ of the single crystal in magnetic fields. 
A sharp peak due to the FIMO is clearly seen in fields between 3.25 T and 6.0 T. 
%Although the peak is not clearly observed at 3 T and 6.25 T, 
A small peak is observed at 3.0 T, and an upturn indicating a peak at a lower temperature at 6.25 T. 
Also, exponential temperature dependences are observed at 2.5 T and 6.5 T, 
indicative of energy gaps for low-lying excitations. 
We conclude from these observations that the critical fields of the single crystal are 
$H_\mathrm{C1}\simeq 3.0$ T and $H_\mathrm{C2}\simeq 6.25$ T. 

A sharp peak in $C(T)$ is clearly observed also in the powder in fields $3.25$ T $\leq H\leq$ 6.0 T as shown in Fig. 2, 
and the peak temperatures are in accordance with those of our previous results \cite{lt23, prl}. 
%(Fig. 1(a)). 
However, the field and temperature dependences of the peak are quite different from those in the single crystal. 
As the field increases from 3.5 T, 
the peak becomes much sharper. 
The field dependence of the peak temperature is weaker than for the single crystal.  
At 3.0 T, 6.25 T, and 6.5 T,  
a sharp upturn is observed indicating a peak at lower temperatures, 
and exponential behaviors are observed at 2.5 T and 6.75 T.   
Based on these features, the gapless field region of the powder is most likely 3.0 T $\leq H\leq$ 6.5 T. 
This field region is wider than that of the single crystal.    
 
Figure 3 is the $H$-$T$ phase diagram of the single crystal and powder obtained from the peaks 
in the specific heat. 
We note two differences between the $H$-$T$ phase boundaries for the two sample forms. 
One is that the peak temperatures are lower for the powder than for the single crystal. 
The other is a difference in the shape of the phase boundary between the FIMO and paramagnetic phase.  
The phase boundary of the single crystal is symmetric with respect to the central field of the gapless field region 
as observed or expected in isotropic spin-gapped compounds investigated so far \cite{FIMOs}. 
In contrast, 
that of the powder is distorted. 
%and the middle field of the gapless phase is slightly higher than that of the single crystals.   
Since the peak in $C(T)$ of the powder is sharp even at 6.0 T in Fig. 1(a), 
it is unlikely that the distinct phase boundary of the powder originates from anisotropy effects. 
In addition, it is revealed by high-field ESR measurements on the powder sample of this compound
that the g-value is almost 2.0 \cite{ESR}.
%The good isotropy of spins in F$_5$PNN is supported by the ESR study on the powder sample of this compound \cite{ESR}; 
%the authors could not use the field marker DPPH which has a g-value of about $g=2.0$ 
%because the ESR absorption line of DPPH overlaps that of F$_5$PNN. 

\begin{figure}[tbp]
   \includegraphics[width=8cm]{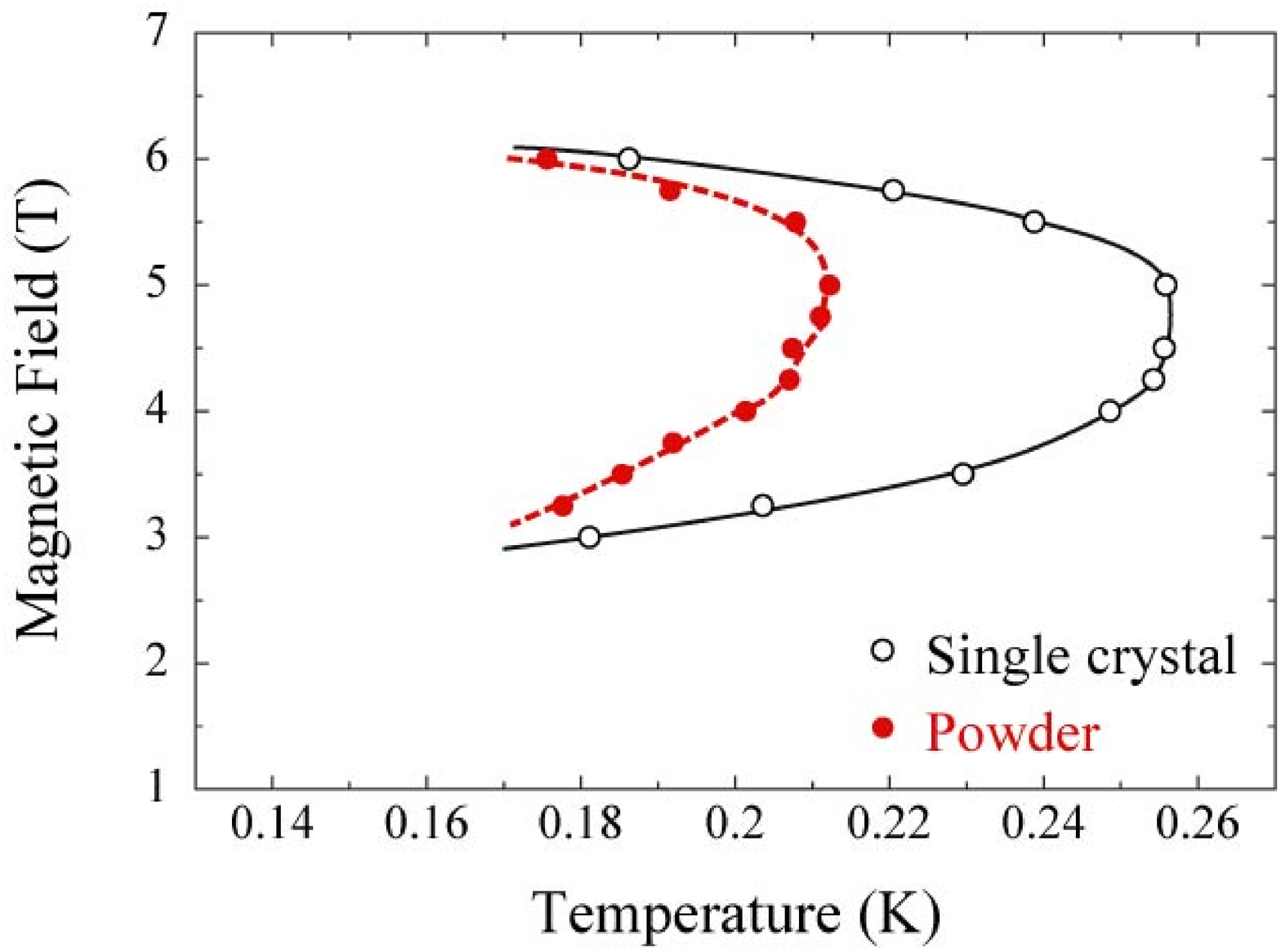}% Here is how to import EPS art
 \caption{(color online) Magnetic field versus temperature phase diagram of the single crystal and powder of F$_5$PNN obtained 
from the specific heat measurements. 
Open and filled circles are peak positions of the specific heats of the single crystal and 
powder, respectively. Solid and broken lines are guides for the eye.}
%and indicating the FIMO phases for single and polycrystalline samples, respectively. 

\end{figure}

\section{DISCUSSION}
The observed distorted phase boundary of the FIMO 
is similar to that of $S=1/2$ strongly frustrated alternating chain models \cite{maeshima}. 
%According to Maeshima {\it et al.} \cite{maeshima}, 
The models exhibit a first-order phase transition at very low temperatures from a conventional 
field-induced antiferromagnetic order of the spin components perpendicular to the external field direction, which is interpreted as 
the BEC of triplet magnons, 
to an IC order along the field direction around the middle of the gapless field region 
where the IC correlation is dominant. 
Because frustration suppresses transverse fluctuations, and then decreases the antiferromagnetic ordering temperature in this field region, 
the phase boundary for the FIMO is distorted. 
To argue the possibility that an IC order is realized in the powder, 
we must first examine if frustration is necessary to explain the powder result. 
%However, in the case of F$_5$PNN, 
%they also suggest that strength of the frustration is not strong enough to have this kind of the incommensurate 
%ordered state \cite{maeshima2}. 

\begin{figure}[t]
\begin{center}
   \includegraphics[width=8.5cm]{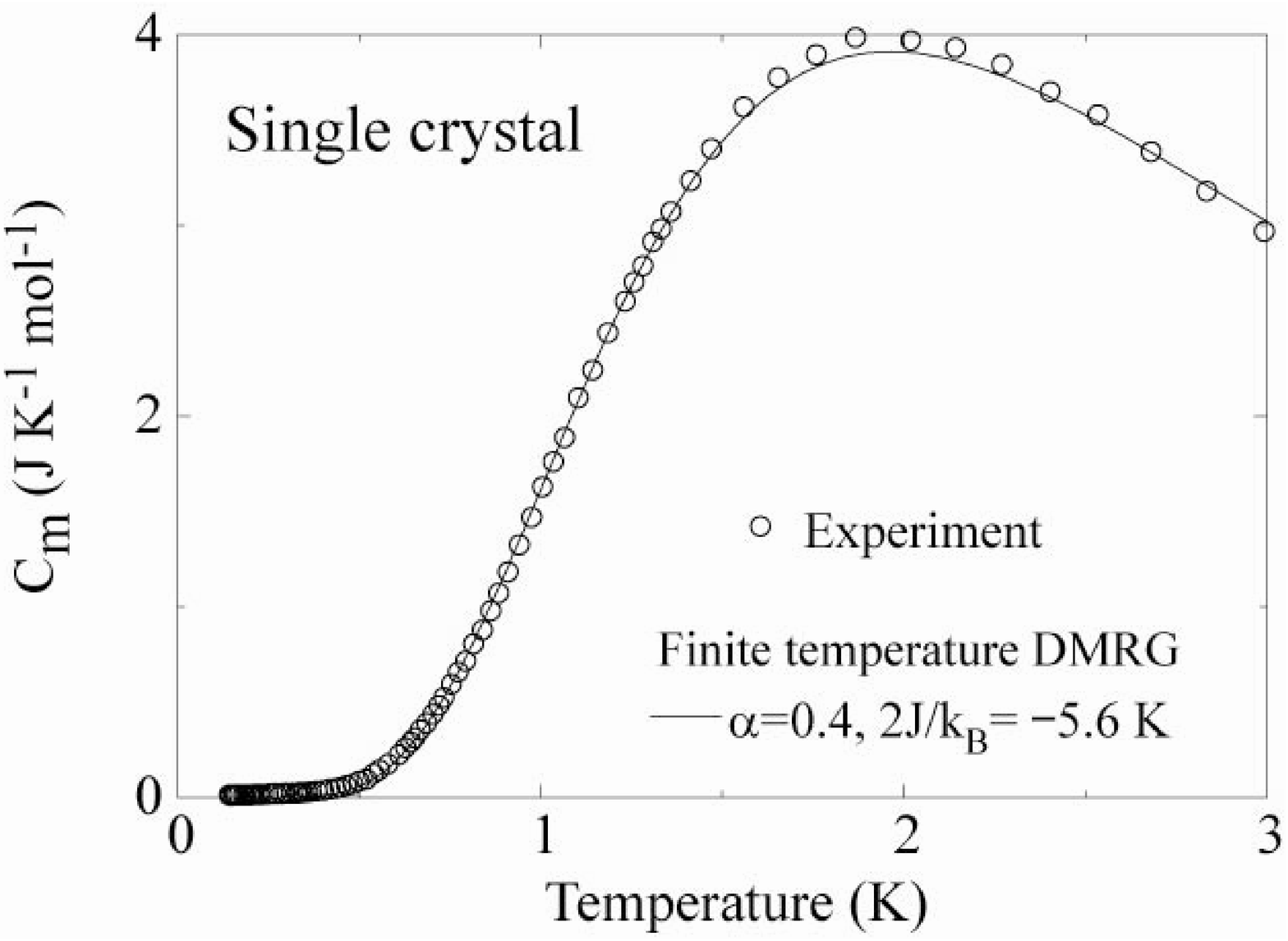}% Here is how to import EPS art
\end{center}
\begin{center}
   \includegraphics[width=8.5cm]{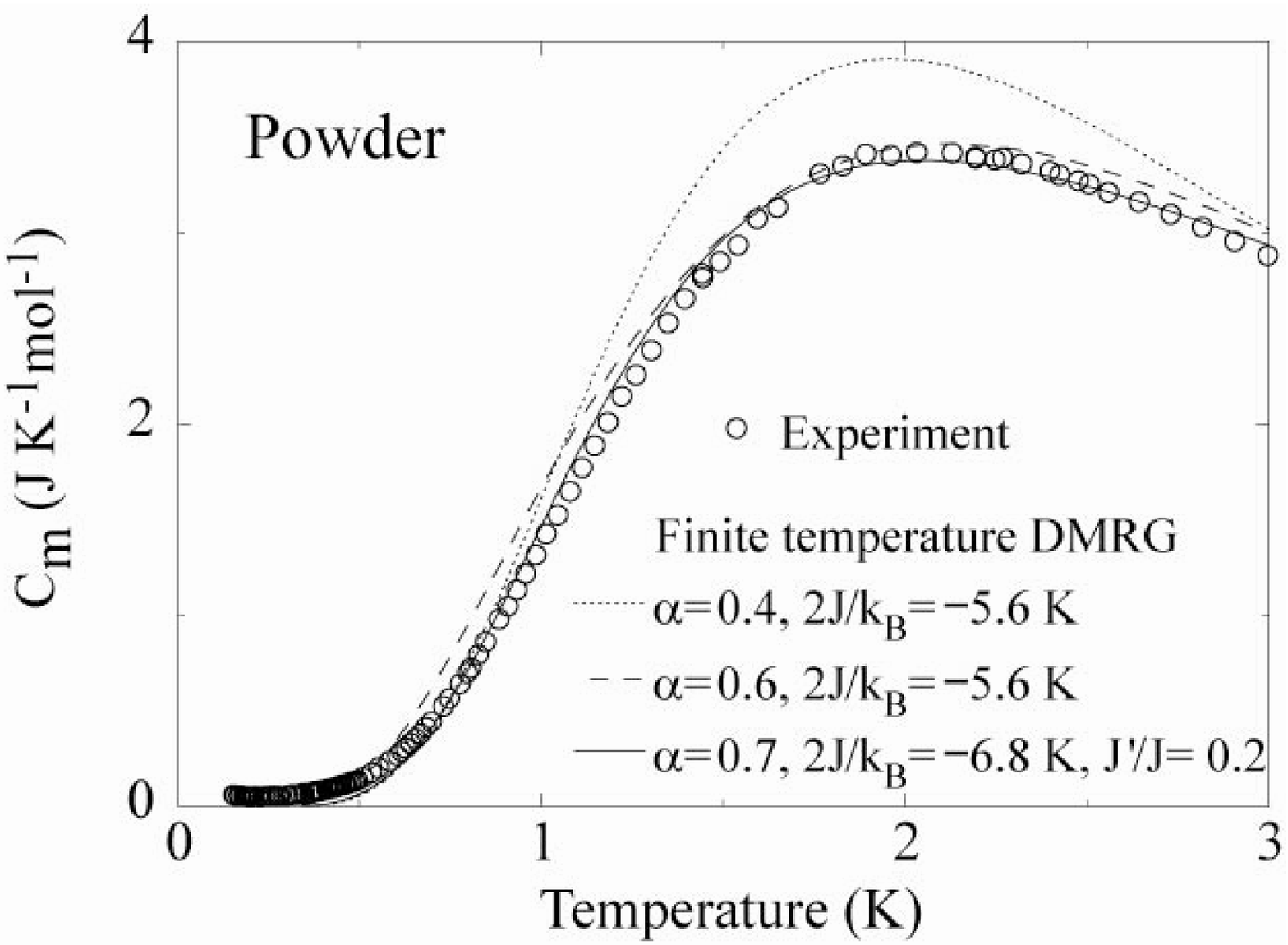}% Here is how to import EPS art
\end{center} 
\caption{Temperature dependence of the magnetic specific heats $C_\mathrm{m}(T)$ of F$_5$PNN single crystal and powder at zero field, 
together with calculated specific heats with three sets of values for the exchange interaction $J/k_\mathrm{B}$ and alternation ratio $\alpha$ 
based on the finite temperature DMRG. 
%The error bar of each experimental point is smaller than the symbol size.
Upper panel: the $C_\mathrm{m}(T)$ of the single crystal and calculation with $2J/k_\mathrm{B}=-5.6$ K and $\alpha=0.4$.
Lower panel: the $C_\mathrm{m}(T)$ of the powder and calculations with $2J/k_\mathrm{B}=-5.6$ K and $\alpha=0.4$, 
$2J/k_\mathrm{B}=-5.6$ K and $\alpha=0.6$, and $2J/k_\mathrm{B}=-6.8$ K, $J'/J=0.2$ and $\alpha=0.7$. 
%Data in fields shown in both glaphs are shifted in the absolute value of specific heat for ease to understand the behaviors. 
%and indicating the FIMO phases for single and polycrystalline samples, respectively. 
}
\end{figure}

To determine the exchange interactions $J$ and alternation ratios $\alpha$ 
of the single crystal and powder, 
we examine the magnetic specific heat at zero field for both samples. 
The lattice contribution to the total specific heat is estimated from the data at zero field so
that the total magnetic entropy for $N$ spins will approach $Nk_\mathrm{B}\mathrm{ln}(2S+1)$ at high temperatures 
where the magnetic susceptibility $\chi $ times temperature $T$ approaches the value for an $S=1/2$ system.
The results are compared with numerical calculations
% with several sets of $J/k_\mathrm{B}$ and $\alpha$ 
based on the finite temperature density matrix renormalization group (DMRG) \cite{DMRG} as shown in Fig. 4. 

The upper panel of Fig. 4 shows the single crystal result and a calculation 
with the set of parameters $2J/k_\mathrm{B}=-5.6$ K and $\alpha=0.4$, 
which have been obtained from the magnetic susceptibility and magnetization of a single crystal \cite{across}. 
The quantitative agreement between the experimental and numerical results 
means that frustration in the single crystal is too small to detect in the specific heat if it exists. 

In the lower panel of Fig. 4, 
we compare the result of the power with numerical calculations
with various parameter sets. 
It should be noted that the calculation with $2J/k_{\rm B}=-5.6$ K and $\alpha=0.4$, 
which well reproduces the single crystal result, 
is largely different from the powder result 
implying the parameters of the powder are not equal to those of the single crystal. 
Although results for $2J/k_{\rm B}=-5.6$ K and $\alpha=0.6$ are better than those for the
first parameter set, clear differences appear in the both side of the peak temperature ($\sim 2$ K). 
Finally, our best result is obtained by assuming a NNN interaction 
for $J/k_{\rm B}=-6.8$ K, $\alpha=0.7$ and $J'/J=0.2$. 
%where the powder has a frustrated interaction and enhanced values of $J$ and $\alpha$.
We note that the enhanced $J$ and $\alpha$ explain the wider gapless field region of the powder 
because $J$ and $\alpha$ govern the width of the gapless field region 
of $S=1/2$ bond-alternating chains \cite{bonner}. 
%thus, this set of the parameters should be valid. 
Also, this agreement rules out the possibility that the distinct phase boundary of the powder is ascribed to 
the disappearance of magnetic moments which comes from the sample deterioration.

\begin{figure}[tbp]
\begin{center}
   \includegraphics[width=8cm]{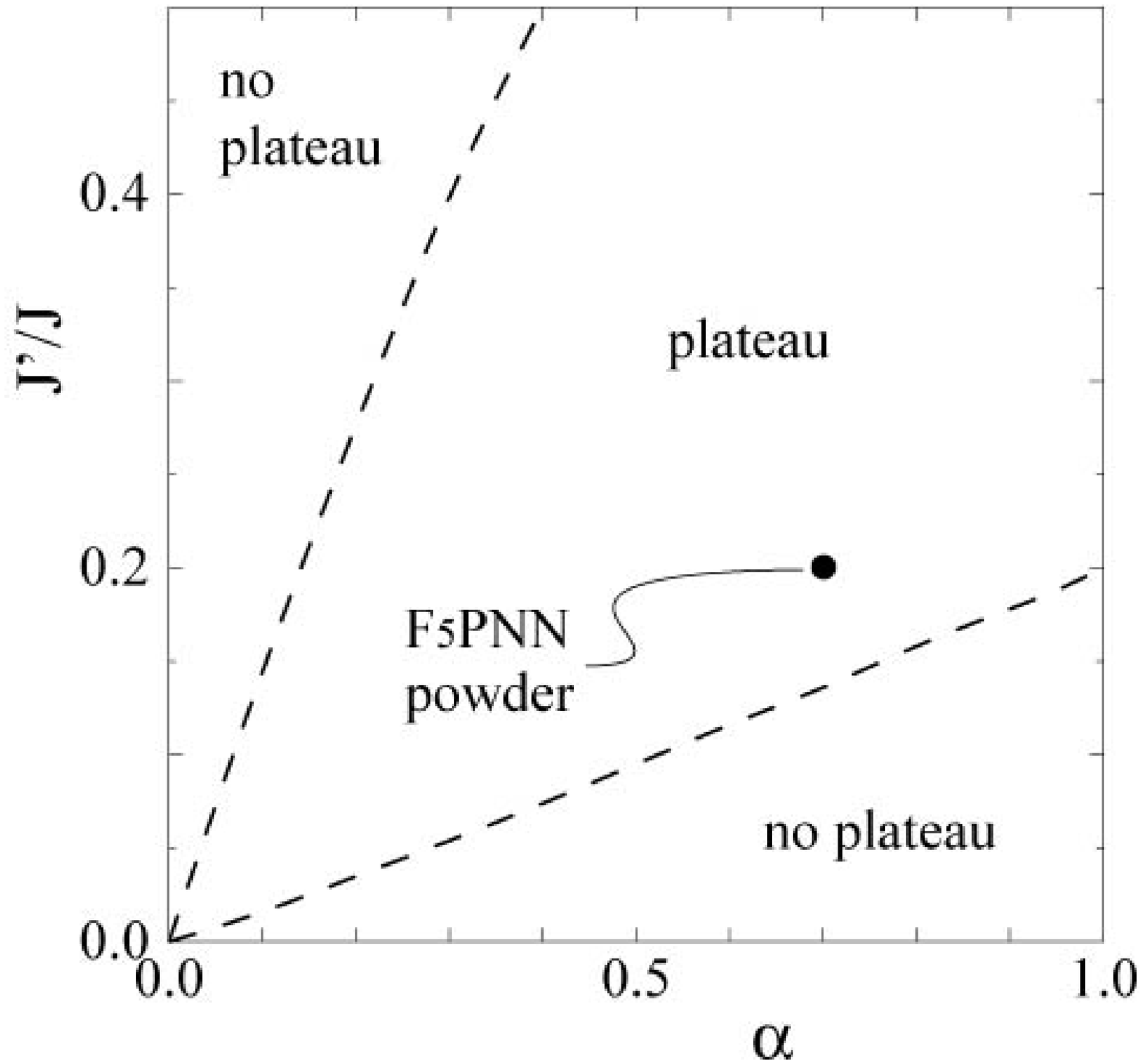}% Here is how to import EPS art
\end{center}
\caption{Alternation ratio $\alpha$ versus NNN interaction $J'/J$ phase diagram at zero temperature 
at the half value of the saturation magnetization, 
which is equivalent to the 1/2 plateau phase diagram in Ref. \onlinecite{tonegawa}. 
The filled circle denotes the set of parameters obtained for the F$_5$PNN powder in this study.}
\end{figure}

The next thing to do is to check whether the set of parameters for the powder is 
comparable to those in which an IC order is theoretically predicted to appear.   
Figure 5 shows the different regions of the dominant correlation at the half value of the saturation magnetization 
in the frustrated alternating chain model as a function of $J'/J$ and the alternation ratio $\alpha$ at $T=0$. 
The IC correlation becomes dominant in the same region where the half-magnetization plateau is stable \cite{maeshima2}. 
The set of parameters for the F$_5$PNN powder, $J'/J=0.2$ and $\alpha=0.7$, turns out to be in this region, 
shown as a filled circle in this figure.  
This result strongly suggests that  the IC correlation is dominant in the powder 
around the center field of the gapless field region and an IC order exists at very low temperatures.  

However, there remains a question why the NNN interaction exists only in the powder.  
The large pressure dependences of the magnetic susceptibility and specific heat of F$_5$PNN 
reported in previous works give us a possible answer to this question \cite{hoso2, mito}. 
According to these works, 
$\alpha $ and $J$ increase with increasing external pressure, and even at $P=0$, 
%At $P=8$ kbar, $\alpha$ becomes 1 from the original value of 0.4 indicating a change from an alternating chain 
%to a uniform chain.   
%Since this pressure is rather small, F$_5$PNN turns out to be very sensitive to an external pressure. 
mixing powder F$_5$PNN with Apiezon N grease changes these values. 
The grease solidifies at low temperatures and gives some stress to the powder inside the solid.   
%Because this compound is highly sensitive to pressure, the influence of stress is not negligible. 
Effective pressure by the solidification of the grease is also reported for the powder of another organic compound \cite{mukai}.   
%In contrast, the larger single crystal may not be influenced by the solidification. 
%The pressure effect may depend on the type of the grease used. 
%In fact, difference is present in the temperature dependence of the magnetic specific heat at zero field between this work and 
%the previous work \cite{mito} in which Apiezon J grease was used instead of Apiezon N.  

Generally, an external pressure enhances inter-chain interactions which increase the ordering temperature of the FIMO.   
Nevertheless, the ordering temperatures of F$_5$PNN is higher for the single crystal than for the powder 
which can be under pressure as mentioned above. 
The strength of an antiferromagnetic interaction in organic magnetic materials depends on how the molecular orbital of an unpaired electron 
overlaps with the others. 
Since this orbital spreads rather widely in each molecule, 
the small variation in the molecular stacking can change the magnetic property drastically \cite{f2pnnno}. 
From this point of view, an external pressure most likely 
changes the molecular stacking in F$_5$PNN  
so that the frustrated NNN interaction, which suppresses the ordering temperature, 
will be enhanced much more than the inter-chain interactions.  
%depending on how molecules stack on one another 
%Therefore, the distorted phase boundaries for the FIMO observed in powder of F$_5$PNN
%may be caused by the frustration enhanced 
%by the slight external pressure or distortion effects to the sample caused in the course of the preparation 
%of the measurements. 
Very recently, we have seen a more clearly distorted phase boundary for the FIMO around the central field 
in the specific heat measurement of deuterated F$_5$PNN powder sample. 
This result will appear somewhere else.  
To investigate quantitatively the pressure-induced frustration in this compound, 
we have proceeded specific heat measurement in magnetic fields under pressure. 
%using a AgCuPd pressure cell \cite{JPSkagoshima}. 

\section{Summary}
We have performed detailed specific heat measurements on the $S=1/2$ alternating chain material F$_5$PNN 
in magnetic fields using a single crystal and powder. 
The shape of the phase boundary for the field-induced magnetic ordered phases is different between the two sample forms. 
We have shown the possibility of the pressure-induced frustration in the powder 
which should lead to field-induced incommensurate ordering around the central field besides 
the Bose-Einstein condensation of triplet magnons, 
by quantitatively comparing zero-field magnetic specific heats of two samples with numerical calculations based on 
the finite temperature density matrix renormalization group. 
A future challenge is the direct observation of the incommensurate ordering.   

%to estimate the exchange parameters.

\section{Acknowledgements}
We thank Yasumasa Takano for helpful advice and valuable discussions. 
We are grateful to Kazuyoshi Takeda, Masaki Mito, 
Seiichiro Suga, Takafumi Suzuki, Akinori Tanaka, Toshihiro Idogaki, Kiyohide Nomura, Luis Balicas, and Takahiro Sakurai for helpful comments. 
Y.Y. was supported by Japan Society for the Promotion of Science.

%\bibliography{apssamp}% Produces the bibliography via BibTeX.

\begin{thebibliography}{8}
%\bibitem[*]{yoshida}Present address: Department of Physics, University of Florida, PO. Box 118440, Gainesville, FL 32611-8440, USA; Electronic address: yoshida@phys.ufl.edu
%\bibitem[$\dagger$]{kawae}Electronic address: yoshida@phys.ufl.edu

%\bibitem{NDMAP} Z. Honda, H. Asakawa, and K. Katsumata, Phys. Rev. Lett., \textbf{93}, 127203 (1998).
%\bibitem{BEC} A. Oosawa, M. Ishii, and H. Tanaka, J. Phys. Condens. Matter, \textbf{11}, 265 (1999). 
\bibitem{nikuni} T. Nikuni, M. Oshikawa, A. Oosawa, H. Tanaka, Phys. Rev. Lett. \textbf{84}, 5868 (2000). 
%\bibitem{nikuni} T. Nikuni, M. Oshikawa, A. Oosawa, and H. Tanaka, Phys. Rev. Lett., \textbf{93}, 127203 (2000). 
\bibitem{suzuki} T. Suzuki and S. I. Suga, Phys. Rev. B. \textbf{70}, 054419 (2004).
\bibitem{maeshima} N. Maeshima, K. Okunishi, K. Okamoto, T. Sakai, Phys. Rev. Lett., \textbf{93}, 127203 (2004); 
J. Phys. Soc. Jpn. \textbf{74}, Suppl. pp. 63 (2005).
\bibitem{maeshima2} N. Maeshima, K. Okunishi, K. Okamoto, T. Sakai, K. Yonemitsu, J. Phys.: Condens. Matter \textbf{18}, 4819 (2006).
%\bibitem{maeshima} N. Maeshima, K. Okunishi, K. Okamoto, and T. Sakai, Phys. Rev. Lett., \textbf{93}, 127203 (2004) ; 
%N. Maeshima, K. Okunishi, K. Okamoto, T. Sakai, and K. Yonemitsu, cond-mat/0604194 (2006).
\bibitem{across} M. Takahashi, Y. Hosokoshi, H. Nakano, T. Goto, M. Takahashi, M. Kinoshita, Mol. Cryst. Liq. Cryst. \textbf{306}, 111 (1997).
%\bibitem{across} M. Takahashi, Y. Hosokoshi, H. Nakano, T. Goto, M. Takahashi, and M. Kinoshita, 
%Mol. Cryst. Liq. Cryst., \textbf{306}, 111 (1997).
\bibitem{izumi} K. Izumi, T. Goto, Y. Hosokoshi, J. P. Boucher, Physica B (Amsterdam) \textbf{323-333}, 1191 (2003).
\bibitem{suga} S. Suga and T. Suzuki, Physica B (Amsterdam) \textbf{346-347}, 55 (2004).
\bibitem{izumithesis} K. Izumi, Master thesis, Kyoto University (2003). 
%\bibitem{izumi} K. Izumi, T. Goto, Y. Hosokoshi, and Jean-Paul. Boucher, Physica B (Amsterdam), 
%\textbf{323-333}, 1191 (2003).
\bibitem{lt23} Y. Yoshida, K. Yurue, M. Mitoh, T. Kawae, Y. Hosokoshi, K. Inoue, M. Kinoshita, K. Takeda, Physica B (Amsterdam) \textbf{323-333}, 979 (2003). 
%\bibitem{lt23} Y. Yoshida, K. Yurue, M. Mitoh, T. Kawae, Y. Hosokoshi, K. Inoue, M. Kinoshita, and K. Takeda, 
%Physica B (Amsterdam), \textbf{323-333}, 979 (2003). 
\bibitem{prl} Y. Yoshida, N. Tateiwa, M. Mito, T. Kawae, K. Takeda, Y. Hosokoshi, K. Inoue, Phys. Rev. Lett. \textbf{94}, 037203 (2005).
\bibitem{wang} X. Wang and L. Yu, Phys. Rev. Lett. \textbf{84}, 5399 (2000).
%\bibitem{prl} Y. Yoshida, N. Tateiwa, M. Mito, T. Kawae, K. Takeda, Y. Hosokoshi, and K. Inoue, 
%Phys. Rev. Lett., \textbf{94}, 037203 (2005).
\bibitem{sample} Y. Hosokoshi, M. Tamura, M. Kinoshita, H. Sawa, R. Kato, Y. Fujiwara, Y. Ueda, J. Mater. Chem. \textbf{4}, 1219 (1994). 
%\bibitem{sample} Y. Hosokoshi, M. Tamura, M. Kinoshita, H. Sawa, R. Kato, Y. Fujiwara and Y. Ueda, 
%J. Mater. Chem. \textbf{4}, 1219 (1994). 
\bibitem{FIMOs} For instance, S. E. Sebastian, N. Harrison, C. D. Batista, L. Balicas, 
M. Jaime, P. A. Sharma, N. Kawashima, I. R. Fisher, Nature \textbf{441}, 617-620 (2006) and references therein. 
%\bibitem{hosoP} Y. Hosokoshi {\it et al.,} Mol. Cryst. Liq. Cryst., \textbf{306}, 423 (1997). 
%\bibitem{hosoP} Y. Hosokoshi, M. Tamura, M. Kinoshita, Mol. Cryst. Liq. Cryst., \textbf{306}, 423 (1997).
\bibitem{ESR} M. Kodama, T. Sakurai, S. Okubo, H. Ohta, Y. Hosokoshi, Proc. 24th Int. Conf. on Low Temp. Phys., pp. 1027-1028 (2006); 
T. Sakurai (private communication).   
\bibitem{DMRG} X. Wang and T. Xiang, Phys. Rev. B \textbf{56}, 5061 (1997); 
N. Shibata, J. Phys. Soc. Jpn \textbf{66}, 2221 (1997).
\bibitem{bonner} J. C. Bonner, S. A. Friedberg, H. Kobayashi, D. L. Meier, H. W. J. Blote, Phys. Rev. B \textbf{27}, 248 (1983).
\bibitem{tonegawa} T. Tonegawa, T. Hikihara, K. Okamoto, M. Kaburagi, Physica B \textbf{294-295}, 39 (2001).
\bibitem{hoso2} Y. Hosokoshi, M. Tamura, M. Kinoshita, Mol. Cryst. Liq. Cryst. Sci. Technol., Sect. A \textbf{306}, 423 (1997). 
\bibitem{mito} M. Mito, T. Kawae, Y. Hosokoshi, K. Inoue, M. Kinoshita, K. Takeda, Solid State Commun. \textbf{111}, 607 (1999).
\bibitem{mukai} K. Mukai, M. Yanagimoto, S. Tanaka, M. Mito, T. Kawae, K. Takeda, Polyhedron \textbf{22}, 2091 (2003).
\bibitem{f2pnnno} Y. Hosokoshi and K. Inoue, Synthetic Metals \textbf{103}, 2323 (1999). 
%\bibitem{mukai} K. Mukai {\it et al.,} Polyhedron, \textbf{22}, 20912098 (2003). 
%\bibitem{JPSkagoshima} K. Yaita {\it et al.} 18pPSB-39, JPS Spring Meeting, 2007. 
\end{thebibliography}

\end{document}